\documentclass[graybox, nosecnum]{svmult}

\usepackage{mathptmx}       
\usepackage{helvet}         
\usepackage{courier}        
\usepackage{type1cm}        
\usepackage{makeidx}         
\usepackage{graphicx}        
\usepackage{multicol}        
\usepackage[bottom]{footmisc}
\usepackage{hyperref}        
\usepackage{soul}            
\usepackage{ulem}
\hypersetup{colorlinks=true,urlcolor=blue}
\usepackage{natbib}
\makeindex             

\def\gsim{\lower.5ex\hbox{$\; \buildrel > \over \sim \;$}}
\def\lsim{\lower.5ex\hbox{$\; \buildrel < \over \sim \;$}}

\begin{document}
\title*{Hard X-ray and gamma-ray detectors}
\author{Vincent Tatischeff\thanks{corresponding author} and Philippe Laurent}
\institute{Vincent Tatischeff \at Universit\'e Paris-Saclay, CNRS/IN2P3, IJCLab, F-91405, Orsay, France, \\ \email{vincent.tatischeff@ijclab.in2p3.fr}
\and Philippe Laurent \at CEA/IRFU/DAp, CEA, CNRS, Université Paris-Saclay, 91191, Gif-sur-Yvette, France, \\ \email{philippe.laurent@cea.fr}}
%
%
\maketitle
{\bf Abstract} 
Space-based astronomy of hard X-rays and gamma rays covers more than seven orders of magnitude in photon energy, from 10 keV to several hundred GeV. Detecting cosmic photons in this energy range is a challenge, due to the relatively low probability of interaction of high-energy photons with matter and the high background noise generated in space detectors by environmental charged particles and radiation. However, the development of new detection technologies is constantly improving the performance of space-based X- and gamma-ray telescopes. This chapter presents the different detectors used in this field of astronomy, their configuration within space telescopes and some proposals for new instruments. 

\vspace{0.3cm}
{\bf Keywords} 
X-ray astronomy; Gamma-ray astronomy; Coded mask; Compton telescope; Pair creation telescopes; Solid-sate detectors; Scintillators; Gas detectors
 
\section{Introduction}
\label{sec:introduction}
The hard X-ray and gamma-ray detectors used in astronomy have often been developed for other fields, including nuclear physics, particle physics, medical imaging and radiation protection. But space detectors must have specific characteristics, particularly in terms of reliability, radiation hardness, compactness, power consumption and operating temperature. The objectives of high-energy space telescopes also differ from those of X- and gamma-ray instruments on Earth, not least the need to detect very low fluxes of astrophysical photons in a background of surrounding energetic particles that can be thousands to millions of times more numerous. 

In this chapter, we first present the different types of telescopes used in high-energy space astronomy and the main associated detectors. We then review the operating principle and characteristics of the various hard X-ray and gamma-ray detectors, followed by a brief presentation of readout electronics and detector performance in orbit. The chapter concludes with an overview of promising technologies for future missions. For more details on the scientific objectives and detection techniques in this field of astronomy, see the recently published comprehensive \textit{Handbook of X-ray and Gamma-ray Astrophysics} \citep{handbook2022}.

\section{Principle of detection of high-energy cosmic photons}
\label{sec:principle}

\begin{figure}[t]
\centering
\includegraphics[width=0.9\textwidth]{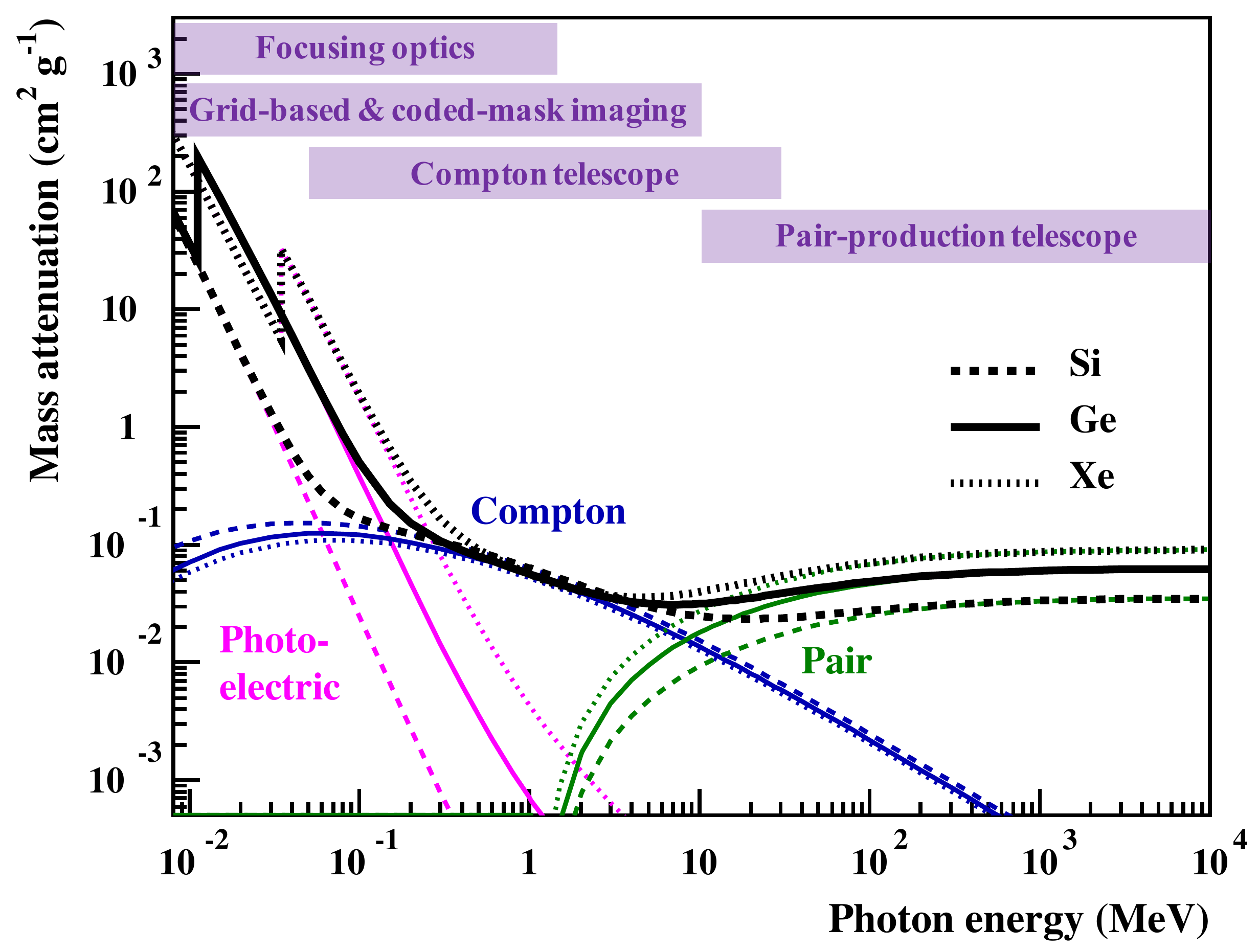}
\caption{Mass attenuation coefficients for (dashed lines) Si, (solid lines) Ge and (dotted lines) Xe, from (magenta) photoelectric absorption, (blue) Compton scattering and (green) pair production (black: total). Data are taken from the NIST XCOM database (\url{https://www.nist.gov/pml/xcom-photon-cross-sections-database}). Also shown are the approximate energy domains for the different types of space telescopes in high-energy astronomy (see text).}
\label{fig:xcom}
\end{figure}

The operating principle of hard X-ray and gamma-ray detectors depends above all on the energy of the photons to be detected. It is based on the three main processes by which photons above 10~keV interact with matter: photoelectric absorption, which prevails in the hard X-ray range, Compton scattering, which is dominant around 1~MeV, and electron-positron pair production, which becomes dominant above about 10~MeV (Figure~\ref{fig:xcom}). The mass attenuation coefficient (i.e. the photon beam attenuation per unit mass of target material), is a strong function of atomic number for photoelectric absorption, but is largely independent of material for Compton scattering (see Figure~\ref{fig:xcom}). Photoelectric absorption is the main interaction process up to 57~keV in Si ($Z = 14$), but up to 150~keV in Ge ($Z = 32$) and up to nearly 300~keV in Xe ($Z = 54$). Pair production becomes dominant over Compton scattering above 15~MeV in Si, 9~MeV in Ge and 6~MeV in Xe (Figure~\ref{fig:xcom}).

\begin{figure}[ht]
\centering
\includegraphics[width=1.0\textwidth]{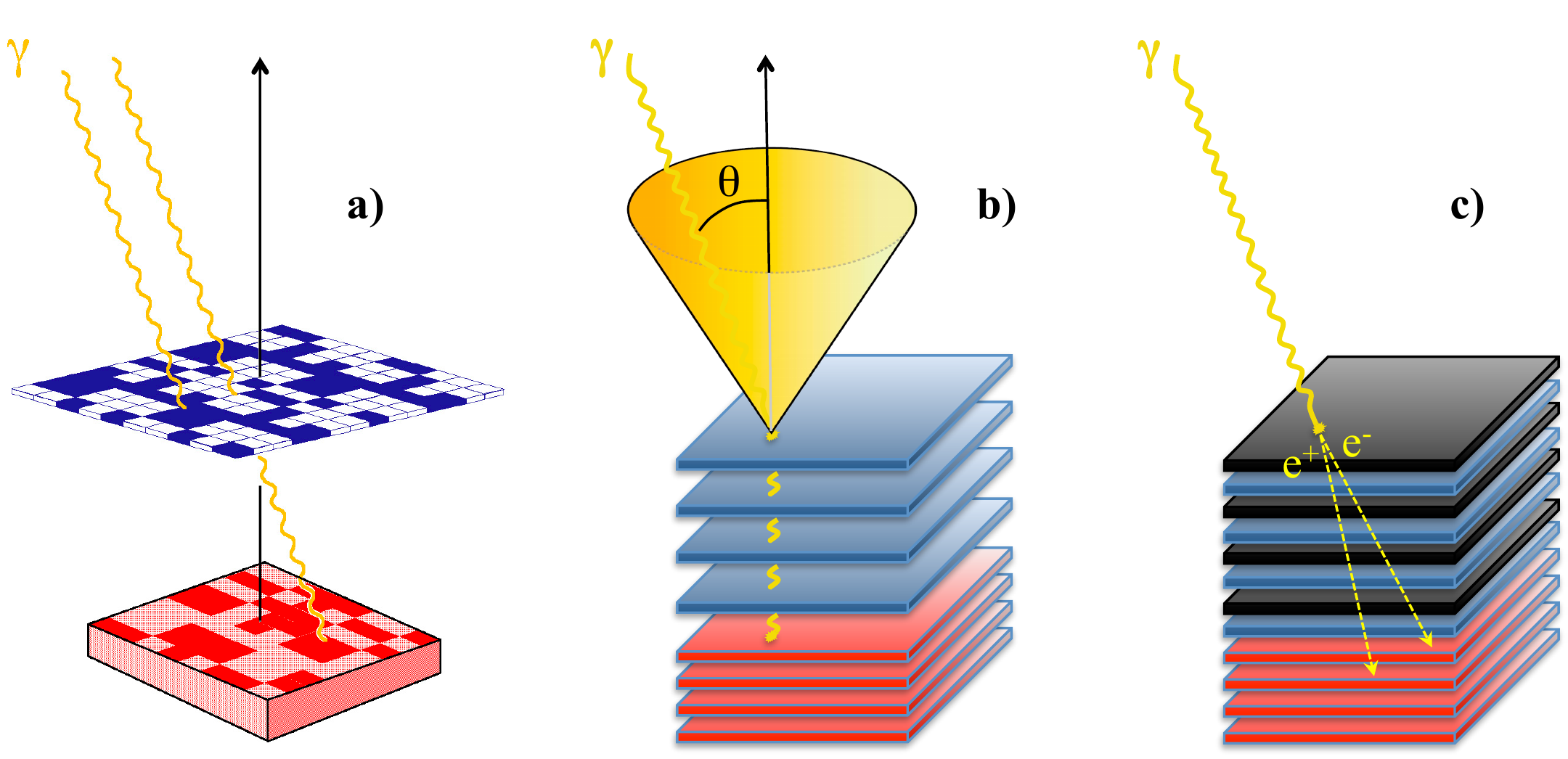}
\caption{Design concepts for (a) a coded-mask telescope, (b) a Compton telescope and (c) a pair production telescope. Here the Compton telescope is made of two detector systems: (blue) a stack of D1 detectors appropriate for Compton scattering and (red) a stack of D2 detectors for efficient absorption of the scattered photons. The pair production telescope in panel (c) consists of (black) layers of high-$Z$ material (e.g. tungsten) to convert incident gamma rays in positron-electron pairs, (blue) position-sensitive detectors to track the $e^+$-$e^-$ pairs and (red) a calorimeter to absorb the secondary particles. Instrument concepts have been proposed which, being devoid of converters, can operate both as a Compton telescope in the MeV range and as a pair-production telescope at higher energies (e.g. \citealt{kanbach2005}).}
\label{fig:concept}
\end{figure}

\subsection{Pixelated detectors for hard X-ray and soft gamma-ray astronomy}
\label{sec:pixelated}
Various imaging systems can be used to focus astrophysical hard X-rays onto a small detection unit, as is classically done in optical telescopes (see Chapter \textcolor{green}{10}, \textcolor{green}{Frontera 2024}). These include grazing incident optics relying on reflection off mirrors in a Wolter-type configuration, as used in the first focusing hard X-ray telescope \mbox{NuSTAR}, which operates from 3 to 79~keV \citep{harrison13}. Focusing telescopes sensitive to higher-energy photons in the hard X-ray and soft gamma-ray band could use diffractive optics systems such as Fresnel or Laue lenses \citep{virgili2022}. With high-spatial-resolution focal plane detectors, such telescopes could achieve sub-arcminute localisation of high-energy sources. Finely-pixelated solid-sate detectors are particularly appropriate for the focal planes of these telescopes. Since the photoelectric cross section is large below $\sim 100$~keV (see Figure~\ref{fig:xcom}), hard X-ray detectors can be quite thin, on the order of a few millimetres. For example, each of two detector units of NuSTAR is comprised of four Cadmium-Zinc-Telluride (Cd(Zn)Te, or CZT) detectors of 20$\times$20~mm$^2$ area and 2~mm thickness that have been gridded into 32$\times$32 pixels. 

Imaging in the hard X-ray and soft gamma-ray domains can also be performed using geometry optics with coded masks or rotating grids (see Chapter \textcolor{green}{13}, \textcolor{green}{Hurford 2024}). The size of the focal-plane detector pixels is also a decisive parameter for the angular resolution of coded-mask instruments, making solid-state detector arrays a good choice. Present and past coded-mask telescopes include IBIS (0.015–10~MeV; \citealp{ubertini2003}) and SPI (0.02–8~MeV; \citealp{vedrenne2003}) onboard the INTEGRAL satellite, Swift-BAT (0.015–0.15~MeV; \citealp{krimm2013}), and the CZT Imager onboard AstroSat (0.02–0.15 MeV; \citealp{bhalerao2017}). Although coded-mask instruments are more appropriate for the sub-MeV range, INTEGRAL SPI and IBIS have imaging capabilities up to several MeV thanks to the use of thick and massive coded-mask elements (e.g. 3~cm thick cells in tungsten alloy for the SPI coded mask). 

\subsection{Detectors for Compton telescopes}
\label{sec:compton}
Using quantum optics in a Compton telescope provides another way of imaging gamma rays in the MeV range (see Chapter \textcolor{green}{11}, \textcolor{green}{McConnell 2024}). This imaging technique does not require an occulting system for the incident gamma-ray flux, which eliminates the need for a large mass of passive materials (i.e. all materials other than detectors). The fraction of active detection mass from the total mass of the instrument can then be maximised. It is based on two successive interactions: Compton scattering in a first detector (often dubbed D1) followed by photoelectric absorption in a second detector (D2; see Figure~\ref{fig:concept}). The photon source can be reconstructed from the Compton formula by measuring the interaction point in three dimensions and energy deposited in each of the detectors. The D1 detector should preferably be made of a material with a low atomic number to favour Compton scattering over photoelectric absorption. In the COMPTEL telescope of the Compton Gamma-Ray Observatory (CGRO), the first Compton telescope to have flown in space, the D1 layer consisted of seven modules of an organic liquid scintillator mainly composed of H and C \citep{schoenfelder1993}. But current proposals for next-generation Compton telescopes favour a stack of Si semiconductor detectors for D1 (see Figure~\ref{fig:concept}), as they can combine fine, sub-millimetre 3-D position resolution and very good spectral resolution \citep{tatischeff2016,caputo2022}. 

The D2 detector (or calorimeter) should be made of a high-$Z$ material for efficient absorption of the scattered gamma-rays, and also present a good energy and 3-D position resolutions for Compton reconstruction. In COMPTEL, the D2 layer consisted of 14 detector modules of NaI(Tl) crystals. Current proposals for a future gamma-ray mission still envisage a ensemble of inorganic scintillators for the calorimeter, but with much finer granularity, e.g. 33,856 bars of CsI(Tl) of 80×5×5~mm$^3$ in the e-ASTROGAM proposal \citep{deangelis17}. Alternatively, the calorimeter can be made up of an array of high-Z semiconductor detectors, such as CdTe or CZT \citep{watanabe2014}.

In the COSI Small Explorer NASA satellite, which is currently planned for launch in 2027, a single type of detector is used for Compton scattering of incident gamma-rays and for absorption of the scattered photons: crossed-strip cryogenic Ge detectors with 3-D position resolution \citep{tomsick2023}. Various other space Compton telescope designs have been tested in the laboratory and by stratospheric balloon experiments \citep[see][]{boggs2006}, but none has been selected by a space agency for a future mission.   
\subsection{Detectors for pair production telescopes}
\label{sec:pair}
Above about 10~MeV, properties of astrophysical gamma-rays are derived from measurements of the electrons and positrons produced by pair creation (see Figure~\ref{fig:xcom}). Space instrumentation for high-energy gamma-rays thus consists of charged particle detectors. The secondary electrons and positrons are tracked with position-sensitive detectors, until they deposit all their energy in a calorimeter. Fine tracking of the $e^+$-$e^-$ pair is required to distinguish incident gamma-rays from the huge charged-particle background in the space environment. 

The first generation of high-energy gamma-ray satellites (SAS-2, COS-B) used spark chambers for particle tracking. A spark chamber consists of a series of conducting layers (typically metal plates or wire grids) stacked in a volume filled with a noble gas (typically neon or argon), where sparks following the particle ionisation path can be measured. A wire-grid, magnetic-core spark chamber interleaved with thin tantalum foils also served as a converter and tracker for the pair production events in CGRO/EGRET \citep[see][]{thompson2022}. In the current generation of high-energy gamma-ray telescopes AGILE \citep{tavani2009} and Fermi/LAT \citep{atwood2009}, tracking gas detectors have been replaced by solid-state detectors: stacks of silicon strip detectors interleaved with tungsten foils for pair conversion (see Figure~\ref{fig:concept}). In both cases, the Si tracker sits on top of a CsI calorimeter to absorb the $e^+$-$e^-$ pairs, and the instrument is surrounded by an anti-coincidence (AC) detector of plastic scintillators to suppress the background induced by cosmic rays. 

Proposals for new high-energy gamma-ray astronomy missions aim to extend the useful energy range for pair production detection downward toward 10 MeV and to develop polarisation. A promising approach could be to use a gaseous time projection chamber as the pair production tracker instead of silicon detectors \citep{hunter2014,gros2018}. 

\section{Detector types}
\label{sec:type}
We will now review the various detectors used in hard X-ray and gamma-ray astronomy, starting with semiconductor detectors, then scintillators, and finally gas and liquid detectors. 

\subsection{Semiconductor detectors}
\label{sec:semiconductor}
Semiconductor detectors enable good efficiency, very good energy resolution and can be manufactured to almost any shape and size. Semiconducting detectors benefit also from a long history of use in particle physics and X-ray and gamma-ray observatories. Among the main semiconductor materials, silicon, germanium and cadmium telluride are the most used currently. Their main properties are given in Table \ref{table:semicon}.

\begin{table}[ht]
\caption{Properties of silicon, germanium and cadmium telluride detectors. A smaller band gap energy ensures a better energy resolution.}
\begin{tabular}{l|c|c|c} 
 \hline
 Semi-conductors & Density & Atomic Number & Gap energy \\ 
                 & (g/cm$^3$) &               &    (eV)    \\
 \hline
  Si	   & 2.33 & 14 & 1.12 \\ 
  Ge     & 5.33 & 32 & 0.67 \\
  CdTe   & 5.85 & 48,52 & 1.44 \\
  Cd(Zn)Te & 5.81 & 48,52 & 1.6 \\
  \hline
\end{tabular}
\label{table:semicon}
\end{table}
The signal induced after a photon interaction into a biased semiconductor is due to the generated charge motion through the sensitive volume, along the applied electric field lines. Charge carrier motion is sensed by capacitive coupling, so that the system geometry defines the charge influence distribution between electrodes. The induced charge on each electrode, given theoretically by the Shockley-Ramo's theorem \citep{shockley38}, is integrated by the corresponding readout preamplifier, during the travel of the charged carriers in the semiconductor.

\subsubsection{Silicon detectors}
\label{sec:silicon}
The long history of silicon use in detectors has led to multiple forms of detector design. The main types of detectors are presented succinctly, since describing all the variations of semiconductor detectors would yield a separate book in itself \citep[see, for instance][]{lutz2007}. Most experiments have very strict specifications, which usually yield a specific design for a specific need.

The most basic silicon detector design that has imaging capacity is a pixel detector. In this case, one side of the detector is polarised at the desired high voltage while the other side is segmented, to be read out by individual electronics. The size of the pixels is not limited by the manufacturing process, but rather by the size of the front-end electronics. One of the main advantages of the pixel detectors is the spectroscopic performance that comes with very small detector sizes. The disadvantage is that the number of necessary channels increases with the square for increasing surface area. This can quickly become a problem when taking into consideration the thermal and mechanical constraints it imposes but also the cost of the detector.

Charge-coupled devices (CCDs) are a very common type of pixelated silicon detector. The difference between a classical pixel detector and a CCD is the particular read-out used. Indeed, CCDs are read out on one side by transferring the charges trapped in individual pixels by a rolling gate voltage applied to the electrodes on one axis. This greatly reduces the number of analogue input channels necessary for the read-out and, as an added advantage, the number of channels scales linearly with increasing surface area.
There are two main limitations of the use of CCDs. First, the limited thickness of the depletion layer, from tens to a few hundreds of micrometers, limits the X-ray detection efficiency. In addition, the readout time of CDDs is necessarily long, as the pixels are read column by column. Added to this is the complexity of the electronics that has to manage the read-out.

\begin{figure}[ht]
\centering
\includegraphics[width=0.7\textwidth]{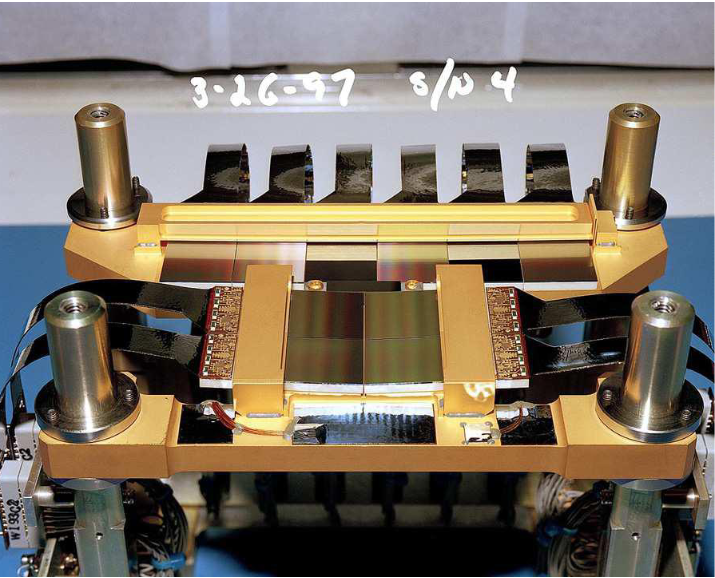}
\caption{View of the ACIS CCD silicon detector onboard the Chandra X-ray Observatory \citep[from][]{Garmire2003}}
\label{fig:CCD}
\end{figure}

Several on-going X-rays space missions are using CCD silicon matrices, such as the ESA XMM Newton EPIC cameras \citep{Briel2000}, the ACIS cameras onboard the NASA Chandra X-ray Observatory (\citealp{Weisskopf2000,Garmire2003}; see Figure~\ref{fig:CCD}), the SXT telescope onboard the Indian ASTROSAT mission \citep{Singh2017} and the Xtend soft X-ray telescope onboard the recently launched XRISM Japanese mission \citep{Hayashida2018,Tashiro2023}.

Active Pixel Sensors (APS) are also pixel devices which have many advantages over CCDs, such as the possibility of implementing electronic functions on-chip, lower power consumption, and high radiation-hardness. They are widely used in particle physics, in the LHC at CERN for instance. They are mainly composed of an active silicon sensor, such as a pin diode, readout by a CMOS pre-amplifier, directly implemented into the pixel matrix in shallow wells. These pixel sensors, also widely used for digital cameras in cell-phones for instance, are used in space for star sensors and alignment systems but have not yet been used in a space application for high energy astrophysics. Regardless, they are considered as an alternative for the double-sided strip detectors envisaged for Compton telescopes (see below), in the framework of the NASA mission AMEGO-X \citep{Fleischhack2022,Suda2024}.

\begin{figure}[ht]
\centering
\includegraphics[width=0.7\textwidth]{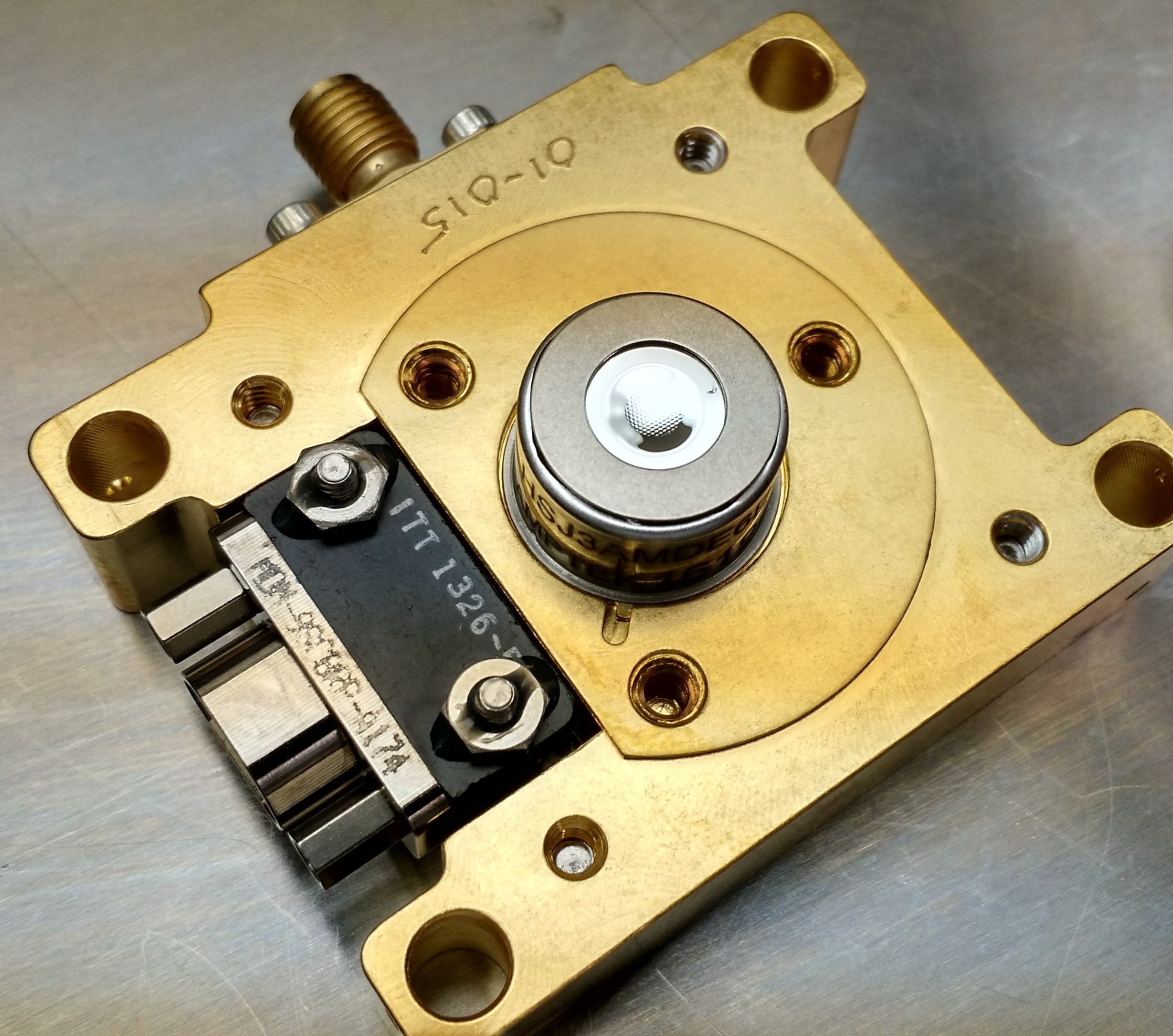}
\caption{View of a NICER single Focal Plane Module (FPM), comprising an SDD detector linked to its CMOS preamplifier in a detector can, seen in the middle of the figure, attached to its gold plated housing (with permission of G. Prigozhin.)}
\label{fig:SDD}
\end{figure}

Silicon drift devices (SDDs) are another variation of a pixel device that are optimised for spectrometry (see Figure~\ref{fig:SDD}). SDDs are read-out at a small anode with charges generated in a bigger pixel volume and then guided by a voltage drift field to the read-out electrode. For radial configuration, a junction field-effect transistor (JFET) can be implemented at the central anode as the first stage of amplification to guarantee fast and low noise readout. The advantage of this type of configuration is indeed the very low noise associated with the anode in comparison to the detector volume. An interesting side-effect of the charge carrier drift is the capacity for sub-pixel position resolution that is achieved
by measuring the drift time. On the other hand, generating the drift voltage gradient
requires more complex polarisation electronics due to the different voltage levels applied.
Currently, SDD detectors are wire bonded, which limits the surface of the detectors to
some maximum wire length, although there are projects that study the feasibility of
bump-bonding SDDs directly to the read-out ASIC similar to pixel detectors.

Instruments using SDD silicon detectors are, for instance, the ``Neutron Star Interior Composition ExploreR'' (NICER) telescope placed on-board the International Space Station since 2017 \citep{Gendreau2016}, or the future Chinese mission eXTP \citep{Santangelo2023}.

Silicon strip detectors are the last type of semi-conductor detector with
imaging capabilities. In at least one detector side, the electrodes span the whole length
of the detector. In the case of single-sided strip detectors (SSD), images are obtained by combining two identical detectors in perpendicular directions $X$ and $Y$. In the case of double sided strip detectors (DSSDs), the electrodes also span the whole length of the detector but are oriented orthogonal to each other on opposite faces. Imaging is achieved by the read-out of both detectors (SSD) or both faces (DSSD) simultaneously, each providing separate axis coordinates. This geometrical configuration greatly reduces the number of read-out
channels necessary when compared to pixel detectors but has the disadvantage of having
worse spectroscopic performance, due to the big size of the electrodes, which increases the input capacitance seen by the front-end electronics. Nevertheless, it can be the only feasible solution when thick and large detectors are needed, as is generally the case for Compton and pair creation space telescopes. For such instruments, detectors as big as $100\times100\times1.5$~mm$^3$ have been manufactured \citep[see, for instance][]{Khalil2016}. 

Single-sided strip detectors are currently used in space for the NASA Fermi LAT telescope, a pair telescope working in the 30 MeV -- 500 GeV energy range \citep{atwood2009, maldera2023}, and composed of 16 towers, each of them consisting of 18 bi-layers (mutually orthogonal) of single-sided strip detectors with interleaved tungsten foils to increase pair creation probability. The trajectories of the electron-positron particles are thus measured by each SSD of the bi-layer, respectively in the $X$ and $Y$ direction.

\begin{figure}[ht]
\centering
\includegraphics[width=0.5\textwidth]{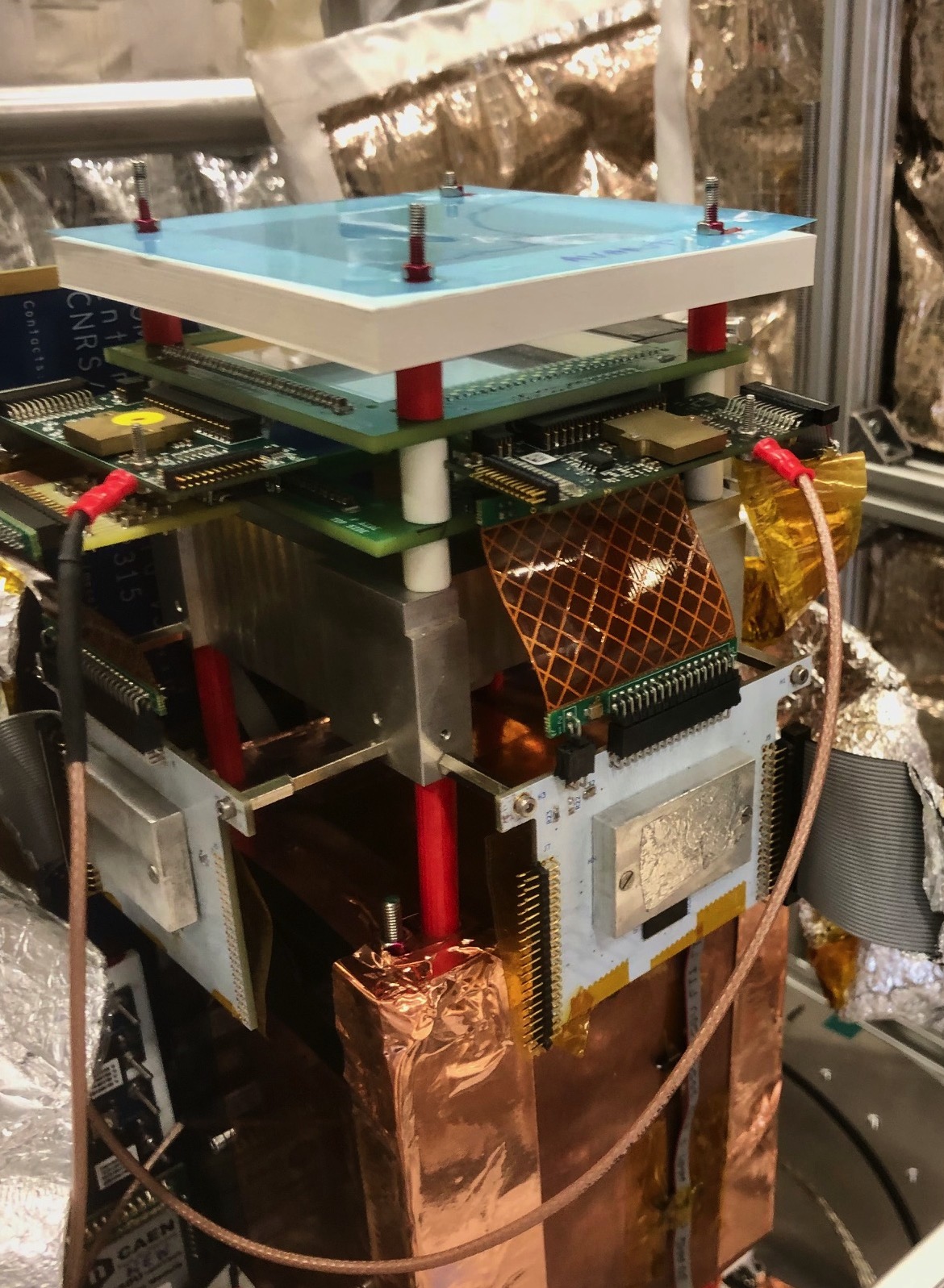}
\caption{View of the COMCUBE polarimeter, prototype of the COMCUBE-S mission. We see on the top, below the cover, two thick DSSD read out by dedicated ASICs.}
\label{fig:DSSD}
\end{figure}

Double-sided Si strip detectors have not yet flown in space but are foreseen for the next generation Compton telescopes such as the ESA ASTROGAM \citep{deangelis17} or the NASA AMEGO projects (however, the new AMEGO-X configuration is based upon active pixel sensors (see above)). DSSDs may also fly before the end of the current decade in the framework of the nanosat swarm mission COMCUBE-S, recently selected by ESA\footnote{See \url{https://www.esa.int/Enabling_Support/Preparing_for_the_Future/Discovery_and_Preparation/One_step_closer_to_a_CubeSat_swarm_mission}}. COMCUBE-S is based upon a constellation of 27 CubeSats, hosting each a Compton polarimeter composed of silicon DSSDs, as well as CeBr3 and GAGG:Ce scintillators (see Figure~\ref{fig:DSSD}).

\subsubsection{Germanium detectors}
\label{sec:germanium}
Germanium detectors are a type of semiconductor well suited to the detection of MeV gamma-rays, as they have a high density and can be produced in large volume. Indeed, high purity germanium (HPGe) detectors can reach diameters of 10~cm for a 20~cm long detector, suitable to detect gamma-rays up to 10~MeV or more.

Due to their low energy gap, germanium detectors are excellent spectrometers, reaching keV resolution for the 1.809~MeV $^{26}$Al astrophysical line for instance. However, they must operate at low temperature at about 80~K, in order to not be blinded by thermal noise. Thus, contrary to silicon or CdTe detectors which can work at room temperatures, germanium detectors have to be cooled, either mechanically by Stirling machines or thermally by plunging them into a liquid nitrogen bath. Also, hole and electron mobilities are of the same order of magnitude, so unlike CdTe detectors (see below the paragraph on CdTe and Cd(Zn)Te detectors), the charge collection time does not vary much for events occurring at different parts of the detector. Germanium detectors are generally read out using electrodes in boron for P-side and lithium for N-side. 

\begin{figure}[ht]
\centering
\includegraphics[width=0.5\textwidth]{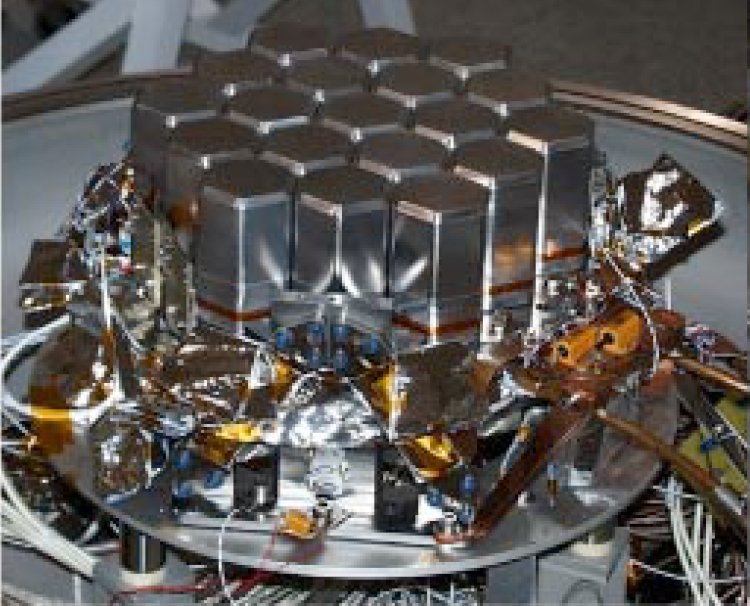}
\caption{View of the 19 germanium detectors inside the INTEGRAL/SPI telescope \citep[from][]{vedrenne2003}.}
\label{fig:SPI_Ge}
\end{figure}

Germanium co-axial detectors have been used in many high energy space missions, such as HEAO-3, RHESSI and INTEGRAL for galactic and solar astronomy observations \citep{Mahoney1980,Lin2002,Roques22}. Germanium detectors excellent energy resolution degrade when the detectors are exposed to particles irradiation. However, the SPI telescope onboard INTEGRAL (see Figure~\ref{fig:SPI_Ge}) has demonstrated that this effect can be, at least partially, recovered by submitting the detectors to an ``annealing", that is a thermal cycle between 80~K and 400~K (see \cite{Roques22} for a review).

Germanium detectors also exist in planar geometries, which were used, for instance for the RHESSI solar mission. These detectors have imaging capabilities, generally implemented thanks to orthogonal strip electrodes, analogous to the silicon DSSD. This kind of detector will be also implemented for the COSI Compton telescope recently accepted by NASA as a Small Explorer mission, for a launch in 2027. Indeed, COSI will use 16 germanium planar DSSDs with 64 strips per side and a 1.162 mm strip pitch \citep{tomsick2023}.

\subsubsection{CdTe and Cd(Zn)Te detectors}
\label{sec:czt}
CdTe shows high efficiency, suitable for detection of photons in the range of 10 to 500~keV, thanks to the relatively high atomic numbers of its constituents ($Z_{Cd}=48$ and $Z_{Te}=52$), which gives a dominant photoelectric absorption probability up to 300~keV against Compton scattering interaction process, and thanks to its high density, around 6~g~cm$^{-3}$. Furthermore, its wide band-gap leads to a very high resistivity and allows operations at room temperature or with moderate cooling ($-20^\circ$~C). Its spectral performance is not as good as germanium and silicon detectors in their own energy range, but is reasonable for most observations of high energy astrophysical sources  (typically 1~keV full width at half maximum (FWHM) at 60 keV and 5~keV at 662 keV). Current manufacturing technologies of CdTe and Cd(Zn)Te allow working with crystal thickness up to a few millimetres for an area in the square centimetre range.

CdTe and Cd(Zn)Te differ by the way they are manufactured. Also, the presence of Zn in Cd(Zn)Te induces a larger band gap and a lower dark current value at room temperature. In the other hand, CdTe is cheaper to produce and can be realised in larger wafers. As for silicon, CdTe semiconductor detectors can exist in single pixels, matrices or strip detectors.

\begin{figure}[ht]
\centering
\includegraphics[width=0.75\textwidth]{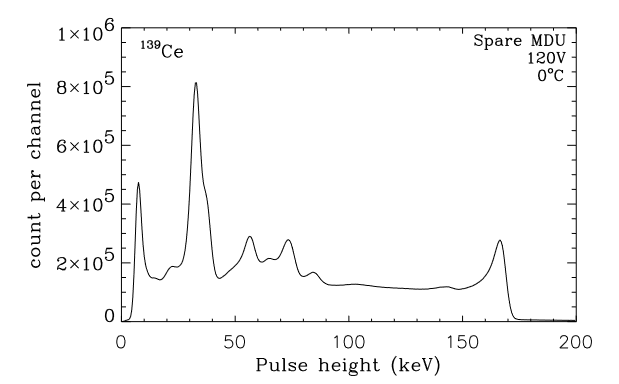}
\caption{Pulse heigth spectrum of a $^{139}$Ce obtained on ground with the IBIS/ISGRI camera. We see clearly the broadening of the low energy part of the $^{139}$Ce 166 keV line ($U = 120$~V and $T = 0^\circ$~C; from \citet{Lebrun2003}).}
\label{fig:Charge loss}
\end{figure}

Photon interactions induce electrons and holes in the sensitive volume, which then induce charges on the electrodes. The carriers transit time depends on their mobility which is low for holes in CdTe semiconductors. Indeed, this can reach several microseconds in the case of a 2~mm thick CdTe under 100~V.  If the peaking time of the readout electronics is faster than this, which is often the case to decrease the impact of the electronics noise, some of the carriers will not be registered. This is called ``ballistic loss". This loss can reach up to 60$\%$ in the case of photon interactions deep in the detector. This loss is visible in CdTe detector spectra, where a broadening in the low energy part of the peaks can be clearly seen (see Figure~\ref{fig:Charge loss}). 

Another loss occurs as charges in the semiconductor can be trapped by the lattice during their transit. This charge trapping defines the lifetime of the different carriers; for electrons and holes in CdTe, this lifetime is of the order of one and ten microseconds for electrons and holes respectively. 

One way to avoid the ballistic loss is to increase the electric field in the semiconductor device. The transit time is then shortened and can be made lower than the electronic peaking time. The caveat is that this bias increase will naturally induce a higher dark current from the detector and thus degrade spectral performances. A process to increase this field, without producing a high dark current, is to use a Schottky indium or aluminium contact on one side of the detector. 

\begin{figure}[ht]
\centering
\includegraphics[width=0.5\textwidth]{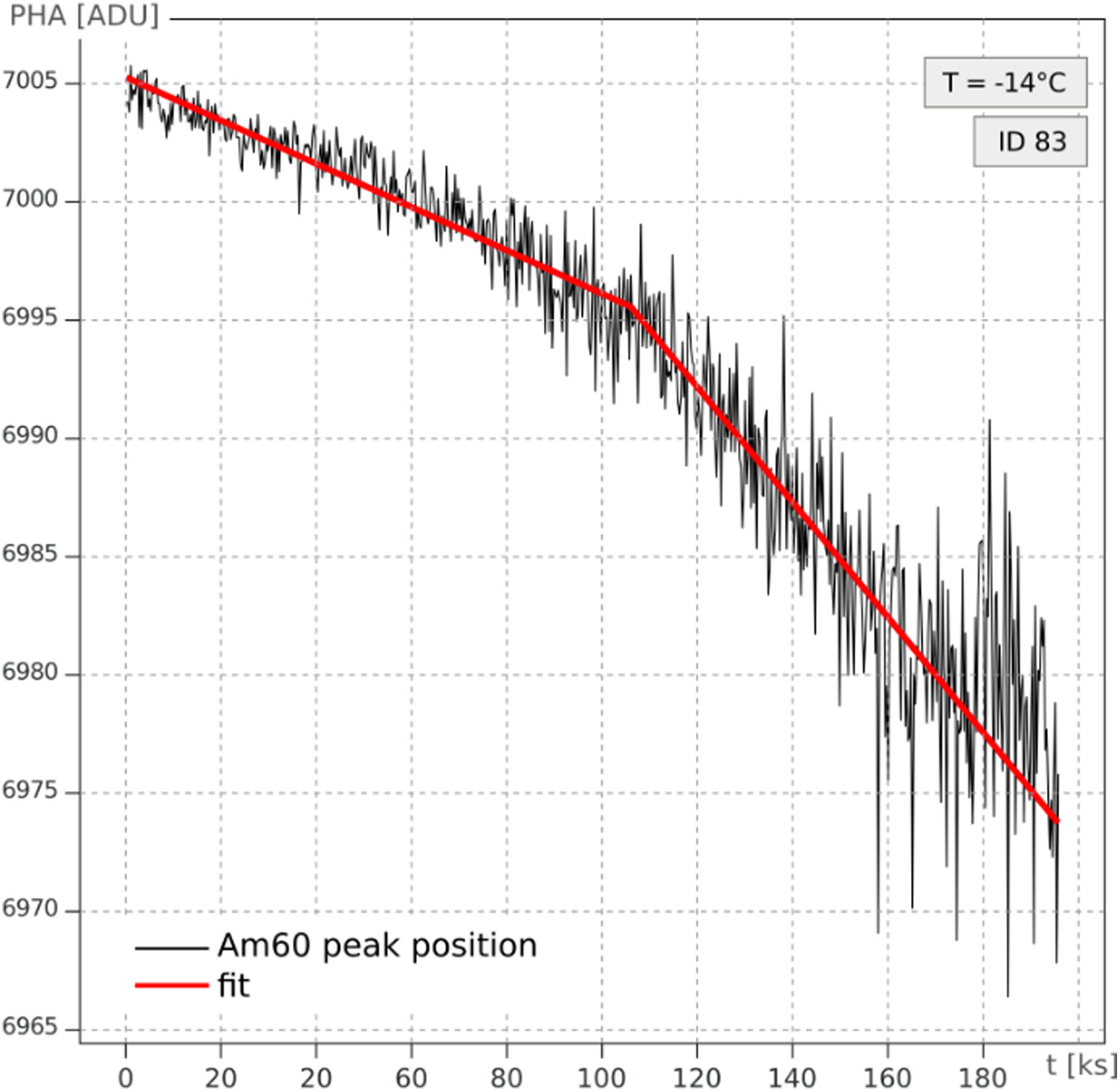}
\caption{60~keV emission line of $^{241}$Am observed with a CdTe detector, similar to those of the HXI instrument on board the Hitomi Japanese mission \citep{Takahashi2018}, during a continuous operation of around 200~ks. We see a clear drop of the detector gain. The red line is a fit which gives a polarisation time of around 107 ks at $U = 250$~V and $T = -14^\circ$~C \citep[from][]{Maier2018}.}
\label{fig:Polarization}
\end{figure}

This ``Schottky barrier" limits the dark current while using higher voltage to bias the CdTe, up to several hundreds of volts. This leads to very good spectral performances (less than 1 keV energy resolution at 60 keV) with a pure Gaussian peak, without any additional broadening. The drawback of this process is the appearance of the ``polarisation effect". Indeed, the detector's gain, as well as the detection efficiency, progressively decreases  with a time constant depending on bias and temperature, until no spectrum can be acquired \citep[see Figure~\ref{fig:Polarization} from][]{Maier2018}. However, the semiconductor retrieves all its properties after a switch off/switch on of the bias voltage. A good way to keep its spectral performances is thus to switch off/on periodically the detectors. This is the strategy applied in the recent space missions using Schottky CdTe detectors.

\begin{figure}[ht]
\centering
\includegraphics[width=0.5\textwidth]{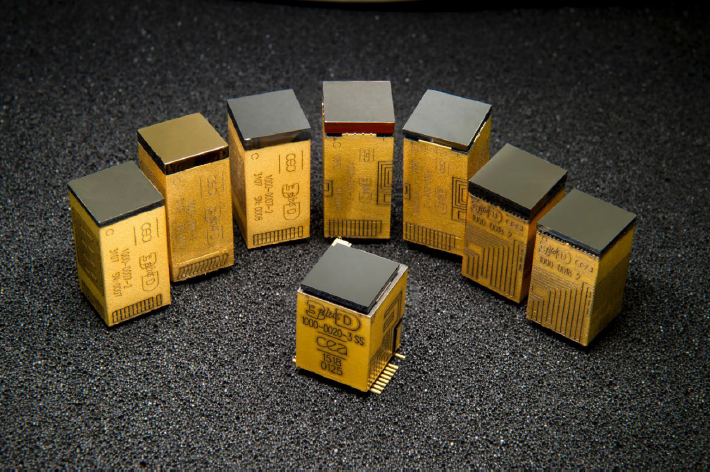}
\caption{The different generations of Caliste CdTe hybrid detectors developed at CEA Saclay. We can see the Caliste-SO at the centre (credit CEA/L. Godart).}
\label{fig:Caliste}
\end{figure}

CdTe detectors have been used in many high energy space missions. Some examples are given below : 
\begin{itemize}
  \item The INTEGRAL ESA mission, launched in 2002, hosts the IBIS coded mask telescope which is composed of two cameras, ISGRI, composed from 16384 4x4 mm CdTe pixels \citep{Lebrun2003}  and PiCsIT, composed of 4096 CsI(Tl) pixels (\cite{DiCocco2003}, see also below in Section \textit{Inorganic scintillators}). These two cameras are still working well after more than 20 years in operation. 
  \item The NASA Swift mission, launched in 2004, hosts the BAT telescope which is also a coded mask telescope based upon Cd(Zn)Te detectors. Its detection plane is composed of 32762 Cd(Zn)Te 2x2 mm pixels \citep{Barthelmy2005}.
  \item Another Cd(Zn)Te camera was developed in the US for the NASA NuStar mission, launched in 2012. The two NuStar separate telescopes have each a focal plane with four 2x2 cm Cd(Zn)Te crystals, divided in 1024 pixels \citep{harrison13}.
  \item The Indian mission ASTROSAT, launched in 2015, hosts the coded mask CZTI telescope. Its detector plane is composed of 16384 Cd(Zn)Te pixels distributed over 64 crystals \citep{bhalerao2017}.
  \item The Japanese Hitomi mission, launched in 2016, used CdTe strip detectors with Shottky barrier (128 strips, pitch 250 µm) for its HXI and SGD telescopes \citep{Takahashi2018, Nakazawa2018}.
  \item DSSD CdTe detectors with Schottky barrier are also used for the seven focal planes of the ART-XC telescope onboard the SRG Russian mission, launched in 2019.
  \item Hybrid 3D CdTe 1 cm$^2$ detectors, called Caliste-SO (see Figure~\ref{fig:Caliste}), are presently used for the ESA Solar Orbiter/STIX telescope \citep{Meuris2012}. 
  \item 6400 Schottky CdTe pixels of 4x4 mm$^2$ and 1 mm  thickness are used for the French-Chinese SVOM/ECLAIRs telescope detection plane, launched in June 2024 \citep{Lacombe2018}.
  \end{itemize}

\subsection{Scintillation detectors}
\label{sec:scintillation}
Scintillation detectors have played a key role in the development of gamma-ray space astronomy. Scintillators can been manufactured in various sizes and shapes at relatively low cost. Coupled to photo-multiplier tubes (PMTs), they have been used since the 1960s on all major gamma-ray space missions \cite[see][and references therein]{iyudin2022}. The first gamma-ray observatory, OSO-1 (Orbiting Solar Observatory 1) carried a hard X-ray telescope operating between 50 and 150~keV, which consisted of a 15~cm$^3$ NaI(Tl) crystal shielded by a lead collimator \citep{peterson1961}. COMPTEL was based entirely on scintillation technologies to explore the sky in the 1~--~30~MeV range \citep{schoenfelder1993}. At higher energies, pair-production telescopes such as CGRO/EGRET \citep{thompson1993}, AGILE \citep{tavani2009} and Fermi/LAT \citep{atwood2009} include a calorimeter made of inorganic scintillators (NaI(Tl) or CsI(Tl)) to absorb the electron-positron pairs and provide a measurement of the incident gamma-ray energy. After more than 60 years of use in space experiments, scintillation detectors are based on a reliable technology that has proved its worth in a wide range of operating conditions. Furthermore, the  development of new scintillating crystals and photo-detectors should renew interest in these detectors for a variety of space applications. 

Scintillators are generally classified as inorganic, organic or gaseous. The latter have a very low scintillation efficiency, typically less than one tenth that of NaI(Tl), which limits their usefulness for a gamma-ray telescope. But combined with a gas proportional counter, they have been used for hard X-ray telescopes \citep[e.g. the High Pressure Gas Scintillation Proportion Counter on-board BeppoSAX;][]{manzo1997}. They are discussed below in Section \textit{Gas and liquid detectors}.

\subsubsection{Inorganic scintillators}
\label{sec:inorganic}
Various inorganic crystals emit scintillation light when charged particles pass through them and deposit  energy through ionisation. In the detection of high-energy photons, the ionisation that causes the scintillation process is due to secondary electrons produced by photoelectric absorption, Compton scattering or pair production. Hundreds of inorganic scintillators have been developed for research\footnote{See the online scintillator library of the Lawrence Berkeley National Laboratory at \url{https://scintillator.lbl.gov/}.}, but only a dozen or so are commonly used as X-ray and gamma-ray detectors. The main inorganic scintillators are alkali-halide crystals to which a doping element (also called ``activator'') is added (e.g. thallium in sodium iodine). The incorporation of an activator impurity into the crystal lattice generates energy states within the forbidden band gap, from which an excited electron (produced by ionisation) can de-excite towards the valence band of the crystal. The scintillation light thus produced can propagate almost freely in the crystal and be collected by a photodetector to be converted into an electrical signal. In the main inorganic scintillators, the light attenuation length, which is defined as the distance the scintillation light must travel to be reduced to 1/e (37\%) of its original intensity, is of the order of 10~cm or more \citep[see, e.g.,][for the optical properties of LaBr$_3$:Ce]{vandam2012}. 

\begin{table}[ht]
\caption{Properties of some common inorganic scintillators. Data are taken from \url{https://scintillator.lbl.gov/} and \cite{knoll2010}, as well as \cite{kamada2012} for GAGG(Ce).}
\begin{tabular}{l|c|c|c|c|c|c} 
 \hline
 Material & Light yield & Peak $\lambda$ & Decay time & Density & FWHM & Hygroscopic \\ 
          & (ph/keV)    & (nm)           & (ns)       & (g/cm$^3$)& @ 662 keV & \\
 \hline
  NaI(Tl)	& 45 & 415 & 230 & 3.7 & 7.1\% & yes \\ 
  CsI(Na) & 43 & 420 & 460, 4180 & 4.5 & 7.4\% & yes \\
  CsI(Tl) & 56 & 550 & 680, 3340 & 4.5 & 5.7\% &	slightly \\
  BGO	    & 10.6 & 480 & 300 & 7.1 & 9.0\% & no \\
  LaBr$_3$(Ce) & 63 & 380 & 16 & 5.1 & 2.9\% & yes \\
  SrI$_2$(Eu) & 120 & 435 & 1200 & 4.6 & 3.0\% & yes \\
  CeBr3 & 58 & 380 & 17 & 5.2 & 4.0\% & yes \\
  GAGG(Ce) & 46 & 520 & 88 & 6.6 & 4.9\% & no \\
  \hline
\end{tabular}
\\
\label{table:inorganic}
\end{table}

The main properties of some common inorganic scintillators are shown in Table~\ref{table:inorganic}. The light yield, i.e. the number of scintillation photons produced by unit energy deposit, affects both the efficiency and the energy resolution of the detector. The latter is presented in Table~\ref{table:inorganic} by the measured FWHM of the $^{137}$Cs gamma-ray line at 662~keV. One of the best scintillation material to date in terms of light output and energy resolution is europium-doped strontium iodide (SrI2:Eu), which could provide an interesting alternative to traditional NaI and CsI scintillators for a gamma-ray space spectrometer \citep{mitchell2019}. However, the quality of a scintillator for spectroscopic measurements also depends on the linearity of its spectral response with deposited energy, which can be affected by various quenching de-excitation processes involving heat generation instead of photon emission \citep{knoll2010}. 

The decay time of the scintillator is an important parameter for the time resolution of detectors, which is particularly relevant for instruments relying on coincidence (or anti-coincidence, AC) measurements between different detectors. New scintillators such as LaBr$_3$:Ce and CeBr$_3$ are interesting in this respect, as well as offering good spectral performance (Table~\ref{table:inorganic}). Another feature to consider for the time response of scintillation detectors is phosphorescence, which can be problematic in many applications. Phosphorescence is a process in which the crystal emits light after scintillation decays, due to the excitation of long-lived electron/hole states in addition to the main scintillation centres. For example, the phosphorescence of CsI(Tl) crystals exposed to cosmic radiation can produce strong pulses lasting several hundred milliseconds \citep{hurley1978}, which can significantly increase the detector background and lead to loss of telemetry \citep{segreto2003}. 

Scintillation detectors are also used to actively shield primary detectors of cosmic hard X-rays and gamma rays from environmental radiation. This is particularly important for collimated or coded mask instruments where the detection of photons coming from outside the instrument field of view must be prevented. The scintillation detector limiting the field of view is then used in an AC mode where an event recorded by the main detector in temporal coincidence with a triggering of the AC detector is rejected. This operating mode is used for example in INTEGRAL SPI \citep{vedrenne2003} and IBIS \citep{ubertini2003}, where the AC detectors of both instruments are made of large crystals of bismuth germanium oxide (BGO). 

The last entry of Table~\ref{table:inorganic} is gadolinium aluminium gallium garnet (Gd$_3$Al$_2$Ga$_3$O$_{12}$) doped with cerium, which is a newly developed scintillator with appealing properties for a gamma-ray instrument: with a light yield of 46 photons~keV$^{-1}$ it is the brightest of the oxide scintillators, its emission spectrum peaking at 520~nm matches well with the sensitivity spectra of silicon photo-detectors (see Sect. \textit{Photo-detectors}), it is a relatively fast scintillator (decay time of 88~ns), has a relatively high density of 6.6~g~cm$^{-3}$ allowing efficient gamma-ray absorption, and is non-hygroscopic, meaning that it does not need to be encapsulated in a hermetically sealed case to avoid contact with humidity. In addition, it shows a good radiation hardness against high-energy proton and gamma-ray irradiation \citep{yoneyama2018}, which is an important characteristic for applications in space missions. GAGG:Ce has already been selected for several nanosatellite missions, including GRID \citep{wen2019}, HERMES \citep{fiore2020}, and CUBES \citep{kushwah21}. 

\subsubsection{Organic scintillators}
\label{sec:organic}
Whereas in inorganic materials, the scintillation mechanism involves the crystal lattice structure, in organic scintillators, light emission comes from transitions between molecular levels \citep{knoll2010}. Organic scintillators can therefore be solid or liquid. Due to the light emission process, organic scintillators are also more prone to quenching than inorganic ones. Quenching, which causes a non-linearity in the scintillator response, is particularly significant for the detection of protons and heavier ions, but less pronounced for gamma rays.   
Organic scintillators generally have densities of around $1$~g~cm$^{-3}$ (see Table~\ref{table:organic}), which are lower than those of inorganic scintillators (Table~\ref{table:inorganic}). For equivalent scintillator dimensions, organic scintillators are therefore less effective at absorbing gamma rays. Moreover, they generally produce less light than inorganic scintillators. Light yields are given in Table~\ref{table:organic} relative to anthracene, which is a common organic scintillator with the chemical formula C$_{14}$H$_{10}$. The light yield of anthracene is $20$~photons~keV$^{-1}$, which is 44\% that of NaI(Tl) (Table~\ref{table:inorganic}).

\begin{table}[ht]
\caption{Properties of some common organic scintillators. Data are taken from \url{https://scintillator.lbl.gov/} and \url{https://detec-rad.com/website/scintillation-materials.html} for p-Terphenyl.}
\begin{tabular}{l|c|c|c|c} 
 \hline
 Material & Light yield & Peak $\lambda$ & Decay time & Density \\ 
          & \% Anthracene & (nm)          & (ns)       & (g/cm$^3$) \\
 \hline
  Anthracene       & 100 & 447 & 30 & 1.25 \\ 
  p-Terphenyl      & 135 & 420 & 3.5 & 1.23 \\ 
  Plastic (BC-400) & 65 & 423 & 2.4 & 1.02 \\ 
  Plastic (BC-422Q)& 11 & 370 & 0.7 & 1.03 \\ 
  Liquid (BC-501A) & 78 & 425 & 3.2 & 0.87 \\  
  \hline
\end{tabular}
\\
\label{table:organic}
\end{table}

Despite these limitations, organic scintillators remain a good choice as scattering detectors in Compton telescopes. In organic materials with low atomic number, gamma rays are more likely to interact by Compton scattering than by photoelectric absorption. The mass attenuation coefficient due to Compton scattering becomes higher than that due to photoelectric absorption above $21$~keV in anthracene, compared to $150$~keV in germanium (see Figure~\ref{fig:xcom}). Thus, the upper detector layer (D1) of the COMPTEL telescope consisted of seven modules filled with NE-213 liquid scintillator (equivalent to the BC-501A scintillator presented in Table~\ref{table:organic}) each read out by an ensemble of eight PMTs \citep{schoenfelder1993}. The pulse shape of the signals from these detectors has been used to distinguish Compton scattering of gamma-ray photons from background events due to inelastic scattering of neutrons produced by cosmic-ray interactions in the atmosphere and the spacecraft. Additional neutron rejection has been obtained by measuring the time of flight between D1 and the second detection plane D2, which consisted of a set of 14 NaI(Tl) detectors for efficient absorption of the scattered gamma rays.

The recent development of new scintillators and photo-detectors can lead to significant improvements in Compton telescopes \citep{mcconnell2016}. For example, the \textit{Advanced Scintillator COmpton Telescope} (ASCOT; \citealp{bloser2018}) is similar in concept to the COMPTEL instrument, with a low-Z organic scintillator for the scatterer (D1) and a high-Z inorganic scintillator for the absorber (D2), but it employs new materials. The low-Z organic scintillator in ASCOT is p-terphenyl \citep{matei2012}, which is a mechanically robust aromatic hydrocarbon of chemical formula C$_{18}$H$_{14}$ producing significantly more light than NE-213/BC-501A (Table~\ref{table:organic}). The high-Z absorber is CeBr$_3$, which is much faster than NaI(Tl) (see Table~\ref{table:inorganic}), allowing time-of-flight measurements between the D1 and D2 layers in a more compact design than COMPTEL. 

Plastic scintillators are materials where luminescent additives are embedded in solid, transparent polymer matrices. One of their major assets is that they are easy to shape into different geometries. Several dozen plastic scintillators have been developed for various applications (see \url{https://scintillator.lbl.gov/organic-scintillator-library/}). BC-400 listed in Table~\ref{table:organic} is one of the plastic scintillators with the highest light output, making it suitable for a wide range of applications, including gamma-ray detection. BC-422Q has a very short decay time of only 0.7~ns and is mainly recommended for ultra-fast timing applications. Thanks to their high efficiency for detecting charged particles and relatively low efficiency for absorbing gamma rays, plastic scintillators can be appropriate for active AC systems aiming at reducing the instrumental background from charged particles in space. Two notable examples are the AC detectors of the gamma-ray telescopes on-board the \textit{AGILE} and \textit{Fermi} missions. They are both made of plastic scintillators covering the top and four lateral sides of the gamma-ray instruments, coupled to PMTs by optical fibers. Both AC detectors achieve an efficiency for charged particle background rejection of the order of 99.99\% \citep{perotti+06,moiseev07}. 

\subsubsection{Photodetectors}
\label{sec:photodetector}

Different types of photodetectors can be used to convert the optical or ultraviolet (UV) light emitted by scintillators into electrical signals that can be processed by an appropriate readout electronics system. Developed in the 1940s \cite[see][]{marshall+48}, PMTs have long been the photo-detectors of choice for space applications. A PMT is composed of a vacuum tube comprising: an input window whose transparency should match the scintillator emission spectrum, a photocathode made of a photo-emissive semiconductor deposited on the inside of the input window, a focusing electrode to accelerate and guide the photo-electron towards the first dynode, an electron multiplier made of a number of dynodes, and an anode collecting the large number of electrons emitted by the multiplier system to produce a sharp current pulse, which is easily detectable. When used in space applications, the mechanical structure of the electrodes has to be reinforced to sustain launch. The quantum efficiency of the photocathode, i.e. the number of emitted electrons per incident photon, now reaches a maximum of about 40\% for a particular wavelength. The overall electron gain of the multiplier typically ranges from 10$^3$ to 10$^8$ depending on the number of dynodes and the inter-dynode potentials set up by the supply voltage. PMTs can be produced in a wide range of sizes, from a few mm to a few tens of cm. 

In recent years, semiconductor photodiode devices have been developed that offer a robust, low-mass, low-volume alternative to PMTs \cite[see][and references therein]{bissaldi22}. Photodiodes also offer the advantages of higher quantum efficiency, of the order of 80\%-90\%, thus potentially better energy resolution, and lower power consumption. A particularly promising photo-detector for scintillator readout in high-energy space telescopes is the silicon drift detector (SDD; see Section \textit{Semiconductor detectors}). It consists in a thin volume of fully-depleted high-resistivity silicon, a few hundred microns thick, with on one side a series of ``drift'' polarising electrodes that generate a strong transverse electric field within the substrate. The electrons produced by X- or gamma-ray interactions in the silicon are then directed by the electric field towards a small collecting anode with a very low capacitance \citep{gatti84}. SDDs have no internal gain, but can have a very low electronics noise. Thus, when coupled to a high light yield scintillator, they can provide state-of-the-art spectral resolution for gamma-ray detection \cite[e.g.][]{fiorini13}. Gamma-ray detectors based on scintillator readout by SDDs are being considered in different space mission projects, including ASTROGAM \citep{deangelis17} and Theseus \citep{labanti20}. 

Photodiodes can be provided with internal amplification by applying a reverse bias voltage to form a high electrical field under the depletion layer \citep{knoll2010}. Charge carriers generated by photo-ionisation are accelerated in the electric field, producing many other electron-hole pairs by the avalanche effect. In avalanche photodiodes (APDs), the required bias voltage is slightly below the breakdown voltage, and the number of created secondary pairs is proportional to the number of absorbed photons. The gain factor is typically of the order of a few hundreds. 

Silicon photomultipliers (SiPMs) are composed of two-dimensional arrays of hundreds or thousands of small (a few tens of microns) APD cells, which are read out in parallel and operated in Geiger mode by applying a bias voltage above the breakdown voltage \cite[see][]{piemonte19}. The gain of an SiPM cell can then reach 10$^5$ to 10$^7$, so that the electrical noise becomes negligible, but the cell output signal is no longer proportional to the number of absorbed photons. However, the summed output signal from an SiPM is proportional to the total number of micro-cells that are triggered by the interactions of optical or UV photons. Thus, when coupled to a scintillator, an SiPM can measure the brightness of scintillation pulses with a good linearity and a gain equivalent to that of a PMT, but with the advantages of being more compact, more robust and requiring no high voltage (typically 20~--~70~V instead of 1~--~3~kV for a PMT). Replacing PMTs by SiPMs in a Compton telescope can be particularly interesting, as it can significantly reduce the mass and volume of passive materials within the instrument and thus increases the detection efficiency \citep{mcconnell2016}. SiPMs are currently one of the most popular photodetectors for scintillator readout. A dozen of space missions using SiPMs have already been launched and several more are under development \citep[see][]{zheng22}.  However, SiPMs are known to be susceptible to radiation damage, their dark current and noise increasing almost proportionally with irradiation fluence above $\sim 10^7$~protons~cm$^{-2}$ or $\sim 10^9$~electrons~cm$^{-2}$ \citep{mitchell21}. The use of this photodetector for a space application may therefore lead to a decrease in mission performance over time, which generally needs to be studied during the mission definition phase. 

\subsection{Gas and liquid detectors}
\label{sec:gas}

\subsubsection{Proportional counters}
\label{sec:proportional}
\begin{figure}[ht]
\centering
\includegraphics[width=0.5\textwidth]{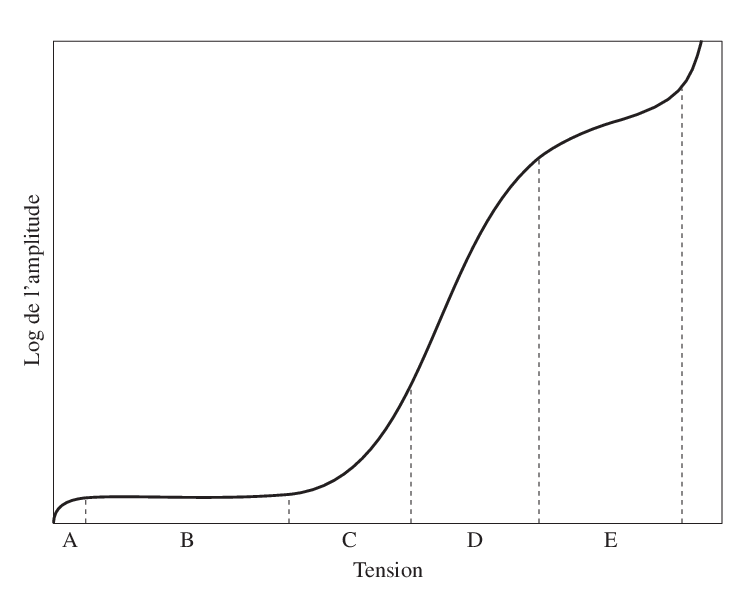}
\caption{Principles of gas detectors. A, B, C, D, and E shows the different regimes linked to the bias applied to the gas chamber. Regime C corresponds to proportional counters, regime E to Geiger-Müller counters. The regime above these corresponds to ``spark chambers".}
\label{fig:gaz}
\end{figure}

The idea of collecting the electric charges induced by the progression of a particle in a gas originated, in fact, at the very beginning of nuclear physics: it was indeed with the help of an ionisation chamber that Pierre and Marie Curie succeeded in proving the existence of natural radioactivity. Gas detectors are all built along the same basic principles: a chamber full of gas with two collecting electrodes placed at each extremity of the chamber, biased at an appropriate voltage. Each time an ionising particle crosses the chamber, it frees electric charges which are collected by the electrodes, which then generate a small measurable pulse. Figure \ref{fig:gaz} shows the evolution of the pulse amplitude with the applied bias for electrons. When the bias is low (regime A in the diagram reproduced in Figure \ref{fig:gaz}), the ionised atoms recombine before the collection of the freed electrons. As the bias gets higher, recombination becomes negligible (regime B). As one increase the bias even higher the freed electrons acquire, in the electric field produced in the chamber, sufficient energy to ionise the gas. This amplification phenomenon tends to increase the pulse amplitude which therefore becomes more easily detectable.
 
As long as the bias is not too high (regime C), the pulse amplitude is proportional, for a given bias, to the ionisation loss of the incoming particle: the  regime of the so-called ``proportional counters". If the electrodes are grids with regularly spaced threading, a knowledge of which thread has been triggered enables the observer to precisely follow the propagation of the charged particle. 
 
As the bias is increased further, the counter leaves the proportional regime (regime D) becoming progressively quasi saturated (regime E): the pulse amplitude then becomes independent of the ionisation losses (this is the Geiger-Müller mode). Finally, if the bias is increased beyond the saturation region, sparks will begin to occur between  the two electrodes; this is the principle behind ``spark chambers". As for proportional counters, the sparks, and thus the interactions of the electron, can be precisely monitored when the electrodes are composed of a network of regularly spaced threads.  

\begin{figure}[ht]
\centering
\includegraphics[width=0.5\textwidth]{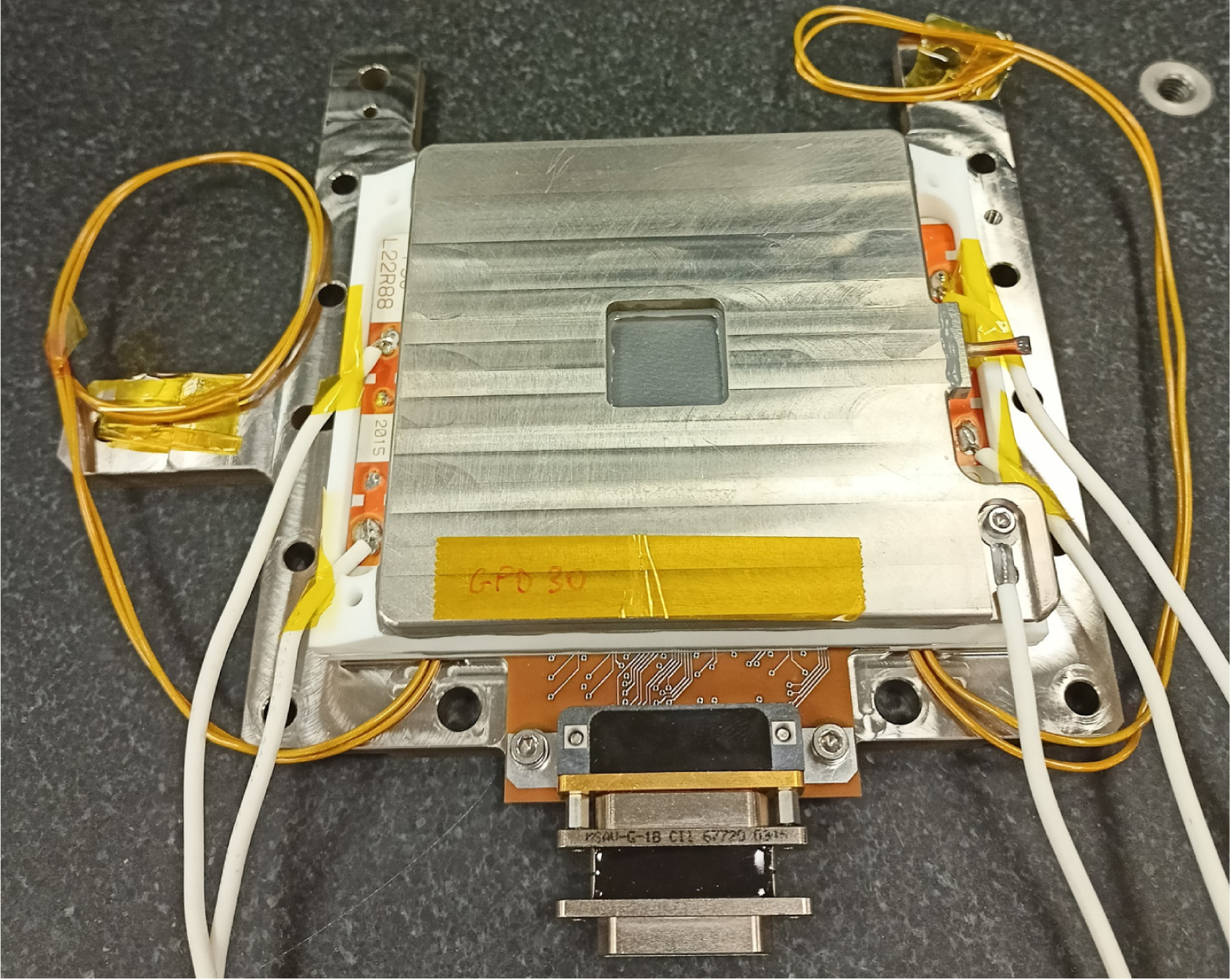}
\caption{View of the IXPE Gas Pixel Detector flight model \citep[from][]{Baldini2021}.}
\label{fig:IXPE}
\end{figure}

Gas detectors are commonly used at both ends of the energy range of gamma-ray space astronomy, even if it is not so easy to use high pressure gas in space. For instance, low energy X-ray telescopes are composed of proportional counters disseminated on a large sensitive area, for which the gamma-ray photons are detected mainly through the photoelectric effect, such as the case of the IXPE space polarimeter (\cite{Soffitta2023}, see also Figure~\ref{fig:IXPE}). To improve the sensitivity of these detectors, the chamber is filled with an high-Z gas at high pressure (up to 5 bars). Another example was the Rossi X-ray Timing Explorer (RXTE), launched in 1995 and stopped in 2012, with its PCA instrument \citep{Jahoda2006}. The Proportional Counter Array (PCA), composed by five gas chambers filled with xenon under one bar pressure, have detected X-ray photons from 2 to 100 keV with $\mu$s accuracy. The JEM-X instrument onboard the ESA INTEGRAL mission \citep{Lund2003}, in operation since 2002, is the first example of a xenon gas detector readout by microstrips in space.

At high energies, spark chambers were commonly used in pair detection telescopes in order to track the propagation of the electron-positron pairs (see Section \textit{Principle of detection of high-energy photons}). They are not used presently for high energy astrophysics, but were widely present in the 80's and 90's, the last example being the EGRET instrument onboard the NASA Compton Gamma-Ray Observatory. The Energetic Gamma-Ray Experiment Telescope (EGRET), detected gamma rays in the 20 MeV -- 30 GeV range, using a spark chamber for direction measurement and a NaI(Tl) scintillator as calorimeter.

\subsubsection{Time projection chambers}
\label{sec:tpc}

A time projection chamber (TPC) is a type of gas or liquid detector in which the trajectory of charged particles can be reconstructed in three dimensions. The active gas or liquid target is immersed in an electric field, where electrons produced along the particle ionisation path drift towards a 2-D position-sensitive electron collection system. The third position coordinate is provided by the measurement of the drift duration. The original TPC design is a cylindrical chamber with multi-wire proportional chambers (MWPC) as endplates. In later designs, the MWPC were replaced by micro-pattern gas detectors (MPGD) such as micromegas, gas electron multiplier (GEM) or micropixels \citep[see][and references therein]{bernard2022}. A major limitation of TPCs in their use for gamma-ray astronomy can be their dependence on information provided by other sub-detectors to build a trigger, in particular to define the start time of the electron drift. 

However, TPCs appear to be promising detectors for improving the sensitivity of space instruments in the energy range where both Compton scattering and pair creation are significant, i.e. between a few MeV and a few tens of MeV (Figure~\ref{fig:xcom}). In a classical Compton telescope, where imaging is based on measurement of the energy and position of the photon scatters in the detectors, the origin of each gamma ray is constrained to an annulus in the sky (called the event circle), which makes it difficult to image an extended source or a field of view containing multiple sources. But using a low-density gas TPC as the D1 scattering detector can enable the scattered Compton electron to be tracked with high precision, thus reducing the event circle to a small arc and improving the instrument point spread function. Prototypes for such an electron tracking Compton camera (ETCC) have been successfully tested in stratospheric balloon flight experiments \citep{takada2022}.

In the low-energy range of the pair production domain, from about 3 to 100~MeV, the performance of current gamma-ray telescopes is strongly limited by their poor angular resolution, which is due to multiple scattering of the $e^+$-$e^-$ pairs in their path through the detectors. The use of a low-density gas TPC as the converter and tracker for the pair production can lead to an improvement of up to an order of magnitude in the single-photon angular resolution, e.g. $\sim 0.5^\circ$ at 100~MeV in the AdEPT \citep{hunter2014} and HARPO \citep{gros2018} TPC prototypes compared to $5^\circ$ at 100~MeV in Fermi/LAT. Despite the lower mass of the active target in a gas TPC compared to that in a silicon tracker like the one of Fermi/LAT, the improved angular resolution could lead to a better detection sensitivity between 3 and 100~MeV. In addition, high-precision tracking of the $e^+$-$e^-$ pairs gives access to the linear polarisation of incident gamma rays, based on measurements of the azimuthal orientation of the pair-creation plane. Linear polarimetry using pair conversion has been successfully demonstrated by the characterisation of the HARPO TPC on a gamma-ray beam \citep{gros2018}. Higher density TPCs using a noble-gas liquid like xenon have also been considered, but the tracking precision was found to be an issue in these devices \citep[see][]{bernard2022}. 

We note that the first gamma-ray TPC in orbit will likely be a ``solid TPC'': in the crossed-strip Ge detectors of the COSI Compton telescope, the third coordinate of the gamma-ray interaction locations is obtained from the difference in arrival time of the electrons and holes at their respective electrodes \citep{tomsick2023}, similarly to a conventional TPC. 

\section{Readout electronics}
\label{sec:electronics}
Gamma-ray sensors have generally two, possibly separated, electronic boards, the Analogue Front-End Electronics (FEE) and the Digital Back-End electronics (BEE). The FEE has several analogue chains which readout the detector's channels. This is done either by using standard electronic components or one or several dedicated ASIC \citep[see, for instance][]{Gevin2021}. In general, the FEE is composed of a charge pre-amplifier followed by a shaper. If the shaper signal passes above a given threshold, the signal is held and the value is passed to the Analogue to Digital converter (ADC) on the BEE board to be digitised. The pre-amplifier gain is tailored to fit the output of the different detectors. The dead-time percentage is dictated by the particle flux incident on the sensor and the acquisition time, which can be fixed or floating, waiting, for instance, that the signal pass below the threshold again.

In the BEE, dedicated rad-hard ADCs are generally used, with a 12-14 bits resolution, depending on the accuracy needed. The read-out sequence is controlled by a micro-controller or an FPGA, which manages the FEE, does some pre-formatting of the data, and sends them to the instrument data processing unit.

The radiation environment in orbit will disturb all equipment which composes a gamma-ray telescope. It can even irreversibly damage the detector media as well as the associated electronic circuits. Degradation of electronic circuits are not specific to gamma-ray telescopes and affect all kinds of space devices whose electronic components can be seriously deteriorated by in-orbit radiation. Such degradation is caused either by the ionisation induced by charged particles or by effects induced by high-energy nuclei: displacements and nuclear spallation.

When an ionising particle crosses an integrated circuit, either a microprocessor or a memory element, it may also release enough charges to change its status, i.e. force a bit 0 to 1, a phenomenon called “bit flip”. The on board software used to manage space equipment may be particularly sensitive to this kind of perturbation, which is generally reversible and can be corrected with dedicated space-proven software. Another effect, the “latch up”, also due to the ionisation of material by charged particles, may induce irreversible effects. Latch up is characterised by the production of a parasitic current along the particle trajectory within a given electronic component. This may result in a short circuit which could cause irreparable damage to the component.
 
Other effects may appear when accelerated particles transfer enough energy to the medium atoms to displace them. This provokes flaws and gaps within the material which may result in various problems depending on the electric component. For example, in the case of a semiconductor, such defects cause electronic noise and even information losses. These defects are more or less persistent, although in some cases thermal cycles are able to attenuate them. Since the probability to create such defects increases when the incident particle energy decreases, one can apprehend the particularly harmful aspect of the intense solar proton fluxes at energies of the order of a few MeV on the instruments electronics.

\section{Detector performances in orbit}
\label{sec:performances}

\subsection{Detector background in orbit}
\label{sec:background}
Satellites in orbit are  bombarded by large fluxes of charged particles, photons and secondary neutrons. These particles and radiation can have two main effects on X- and gamma-ray instruments: (i) radiation damage from high-energy particles interacting with the detectors and their readout electronics, and (ii) degradation of the instrument sensitivity. Charged particles can alter the state of electronic integrated circuit components, causing electronic noise and errors such as single-event upset and latch-up (see Section \textit{Readout electronics}). To avoid such damage and malfunctions, space electronics use radiation-hardened components that are resistant to the effects of ionising radiation. High-energy particle irradiation can also lead to an increase of the leakage current and noise of solid-state detectors, which can result in a loss of instrument performance in the course of the mission \citep[see, e.g.,][for the radiation damage assessment of SiPMs]{mitchell21}. 

\begin{figure}[t]
\centering
\includegraphics[width=0.99\textwidth]{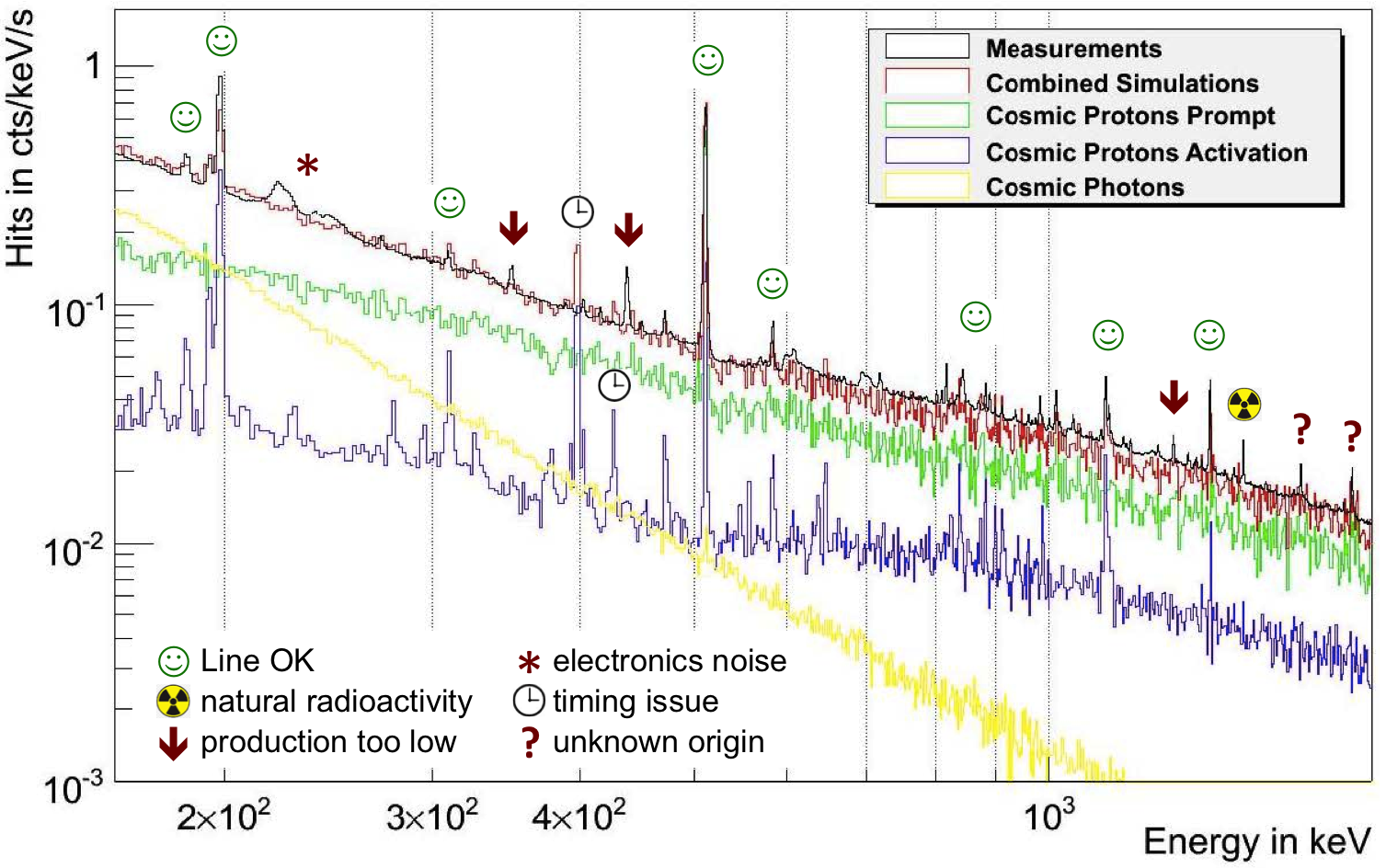}
\caption{Comparison of a background spectrum from the \textit{Wind}/TGRS instrument (black) with the main components simulated using MEGAlib software. Most of the lines are well explained by the activation of the TGRS Ge detector and \textit{Wind} spacecraft by cosmic protons (blue). The broad features in the data between 210 and 260 keV are electronic artefacts. The 1461 keV line comes from the natural radioactivity of $^{40}$K ($T_{1/2}=1.25 \times 10^9$~y), which is not taken into account in the simulation. Adapted from a presentation by A. Zoglauer on the MEGAlib website (\url{https://megalibtoolkit.com/}); see also \cite{weidenspointner2005}.}
\label{fig:megalib}
\end{figure}

High-energy particles and environmental radiation can trigger space detectors and thus generate bad data contributing to the background noise. The flux of energetic particles in space is generally thousands to millions of times greater than the weak flux of gamma-ray photons from the astrophysical sources of interest \citep[see][]{tatischeff2022}. This background noise can be partly suppressed by shielding the gamma-ray instruments with an anti-coincidence detector highly sensitive to charged particles, but less sensitive to gamma rays \citep[e.g.][]{moiseev07}. However, cosmic rays also produce radioactive nuclei by spallation reactions in the spacecraft, and these radioactivities can induce a delayed background in the detectors that is more difficult to reduce (see Figure~\ref{fig:megalib}).  

The choice of orbit can be critical to the performance of a high-energy mission. In terms of background count rate, the best environmental conditions are found in near equatorial, low-Earth orbits \citep{tatischeff2022}. However, detectors in such an orbit are subject to continuous day-night variations, which are accompanied by potentially problematic temperature changes. A satellite in high-Earth orbit is exposed to a higher particle background, but can benefit from a more stable environment, except during periods of intense solar activity. 

\subsection{Detector response simulation}
\label{sec:simulation}
Numerical simulations have become essential for designing new X- and gamma-ray space telescopes and determining their response function while in operation. The ever-increasing complexity of the instruments and the specific environmental conditions of satellites in orbit make it impossible to estimate accurately the performance of a high-energy mission without Monte Carlo simulations. These can include the detection of high-energy photons from cosmic sources as a function of the source position and energy spectrum, the interaction of charged background particles with the instrument and spacecraft, as well as the response of the electronics to energy deposition in the active volumes of the detectors. Increasingly accurate numerical models have been developed to describe the environmental background components as a function of orbit \citep[see, e.g.,][]{cumani2019}. 

The standard Monte Carlo toolkit used in high-energy physics to simulate the passage of particles and photons through matter is Geant4 \citep{allison2016}. Several software packages using Geant4 have been developed for specific space applications, like MEGAlib designed mainly for Compton telescopes \citep{zoglauer2006}, and BoGEMMS for pair-tracking telescopes \citep{aboudan2022}. These tools have proved highly effective for predicting the performance of new missions and for interpreting space data (see Figure~\ref{fig:megalib}). Specific software have also been developed to analyse data from particular space instruments, such as Fermitools for data produced by the Fermi Gamma-Ray Space Telescope (\url{https://github.com/fermi-lat/Fermitools-conda/}).

\section{Outlook}
\label{sec:outlook}
Hard X-ray and gamma-ray astronomy has made exceptional progress since the first cosmic gamma-ray sources were detected in the 1960s. Hard X-ray survey catalogues now include more than a thousand objects detected below 200~keV \citep{oh2018,krivonos2022} and the latest Fermi/LAT catalogue contains 6658 sources in the energy range from 50~MeV to 1~TeV \citep{abdollahi2022}. New transient gamma-ray sources, mostly gamma-ray bursts, are now detected every day \citep{vonkienlin2020}. However, the so-called ``MeV domain" between about 200~keV and 50~MeV remains largely unexplored, due to the lack of sensitivity of past and current missions operating in this energy range (see Chapter \textcolor{green}{3}, \textcolor{green}{Kanbach \& Greiner 2024}). The upcoming COSI Small Explorer NASA satellite will provide new data in this energy range \citep{tomsick2023}, but further reducing the ``MeV sensitivity gap" is currently a very active field of conceptual and technological development \citep[see, e.g.,][]{siegert2022}. 

New diffraction-based gamma-ray optics, such as Laue or Fresnel lenses, are now making it possible to develop a focusing telescope sensitive beyond 200~keV (see Chapter \textcolor{green}{10}, \textcolor{green}{Frontera 2024}). A Compton telescope based on modern detector technologies could also achieve a significant gain in sensitivity in the MeV range, by about two orders of magnitude over CGRO/COMPTEL (see Chapter \textcolor{green}{11}, \textcolor{green}{McConnell 2024}). The development of large-area Si DSSDs readout by state-of-art integrated electronics circuits makes it possible to build a Si tracker with excellent energy and position resolution for the scatterer detector of an advanced Compton telescope \citep{kirschner2024}. Alternatively, pixelated Si sensors developed for particle physics seem promising for gamma-ray space applications as well \citep{steinhebel2024}. New scintillators such as GAGG(Ce) coupled to solid-state photodetectors such as SiPMs \citep[e.g.][]{kushwah21} can offer a high stopping power together with good spectral and spatial resolutions for the calorimeter of a future Compton telescope. Alternatively, an array of state-of-the-art semiconductors such as Cd(Zn)Te or CZT can also provide a very accurate measurement of the interaction location and energy deposit of the scattered Compton photons \citep{meuris2022}. TPCs too appear to be promising detectors for improving the sensitivity of space instruments  between a few MeV and a few tens of MeV \citep{bernard2022}. Thus, several detection technologies are now available to improve significantly the performance of next-generation gamma-ray space telescopes.

\section{Cross-References}

\begin{itemize}
    \item  \textcolor{blue}{Gamma-ray astronomy}
    \item  \textcolor{blue}{Hard X-ray/soft gamma-ray Laue lenses for high energy astrophysics}
    \item  \textcolor{blue}{Compton scattering for imaging and polarimetry}
    \item  \textcolor{blue}{X-ray imaging with masks and grids}
\end{itemize}

\begin{acknowledgement}
The research presented in this chapter has received funding from the European Union’s Horizon 2020 Programme under the AHEAD2020 project (grant agreement n. 871158). We thank Aline Meuris and Joseph Mangan for their careful reading of the manuscript.
\end{acknowledgement}

\bibliography{references}
\end{document}